# On Modelling and Prediction of Total CPU Usage for Applications in MapReduce Enviornments

*Nikzad Babaii Rizvandi*[1,2], *Javid Taheri*[1], *Reza Moraveji*[1,2], *Albert Y. Zomaya* [1]

[1] Centre for Distributed and High Performance Computing
School of IT, University of Sydney, Australia

[2] National ICT Australia (NICTA), Australian Technology Park

nikzad@it.usyd.edu.au

***Abstract***— recently, businesses have started using MapReduce as a popular computation framework for processing large amount of data, such as spam detection, and different data mining tasks, in both public and private clouds. Two of the challenging questions in such environments are (1) choosing suitable values for MapReduce configuration parameters –e.g., number of mappers, number of reducers, and DFS block size–, and (2) predicting the amount of resources that a user should lease from the service provider. Currently, the tasks of both choosing configuration parameters and estimating required resources are solely the users' responsibilities. In this paper, we present an approach to provision the total CPU usage in clock cycles of jobs in MapReduce environment. For a MapReduce job, a profile of total CPU usage in clock cycles is built from the job past executions with different values of two configuration parameters e.g., number of mappers, and number of reducers. Then, a polynomial regression is used to model the relation between these configuration parameters and total CPU usage in clock cycles of the job. We also briefly study the influence of input data scaling on measured total CPU usage in clock cycles. This derived model along with the scaling result can then be used to provision the total CPU usage in clock cycles of the same jobs with different input data size. We validate the accuracy of our models using three realistic applications (WordCount, Exim MainLog parsing, and TeraSort). Results show that the predicted total CPU usage in clock cycles of generated resource provisioning options are less than 8% of the measured total CPU usage in clock cycles in our 20-node virtual Hadoop cluster.

## 1. Introduction

Recently, businesses have started using MapReduce as a popular computation framework for processing large amount of data in both public and private clouds; e.g., many web-based service providers like Facebook is already utilizing MapReduce to analyse its core business as well as to extract information from their produced data. Therefore, understanding performance characteristics in MapReduce-style computations brings significant benefit to application developers in terms of improving application performance and resource utilization.

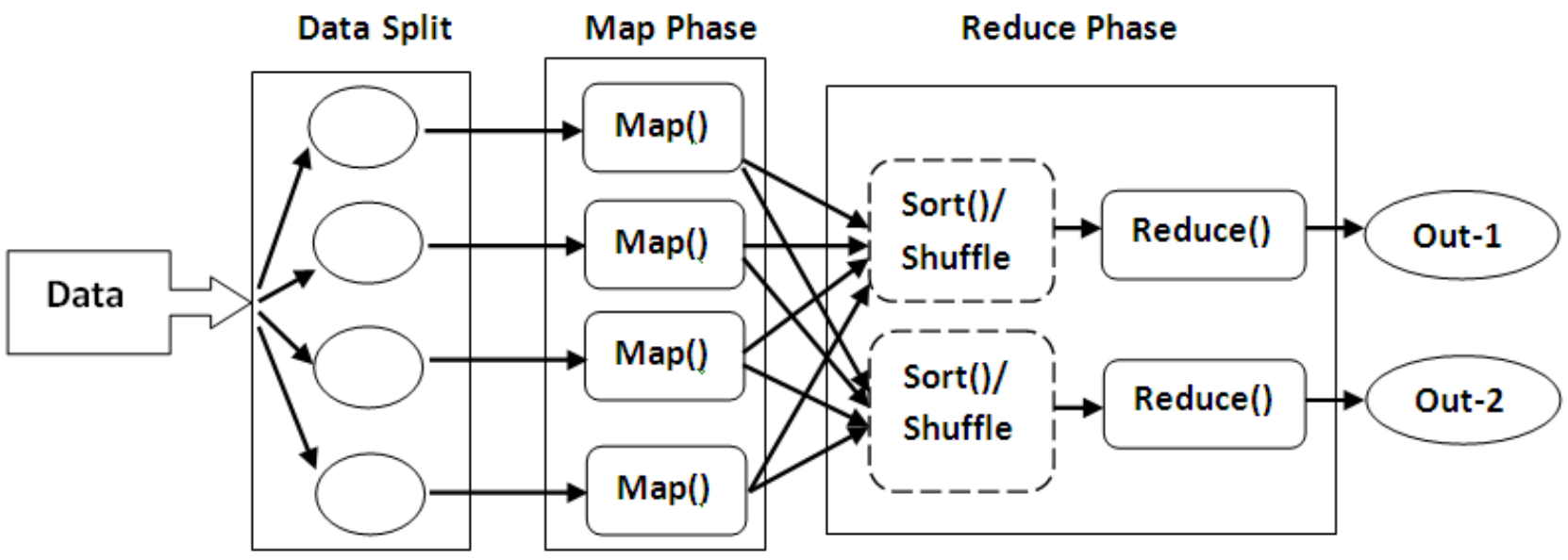

*Figure 1.MapReduce workflow*

One of the regular user jobs –running experiments on MapReduce environment– is to frequently process and analysis almost relatively fixed-size data. For example, system administrators are always interested to frequently analysis system log files (such as Exim MainLog files[1]). As these log files are captured with fix sampling rate, their sizes do not usually change for specific period of times –e.g., for each month. Another example is Seismic imaging data where fix number of ultrasound senders/receivers produce earth underground information in a specific region; therefore, the size of output file –usually in the order of terabyte– is usually consistent [2]. The other example is to find a sequence matching between a new RNA and RNAs in a database [3], where the size of such databases (such as NCBI [4]) is almost unchanged over adjacent periods of time. These applications, which generally heavily consume resources, repeatedly show same execution pattern over their frequently deployments. As a result, any improvement in their resource utilisation can significantly improve the overall performance of such systems

Two typical performance questions in MapReduce environments are: (1) how to estimate the required resources for a job, and (2) how to automatically tweak/tune MapReduce configuration parameters to improve execution of a job; these two questions are important as they directly influence the performance of MapReduce jobs. Moreover, users are solely responsible to properly set these configuration parameters to achieve desirable performances. Although there are a few recent methodologies to estimate resource provisioning of MapReduce jobs (mostly on execution time prediction [5-8]), to best of our knowledge, there is no practice to study the dependency between performance of executing a job and the configuration parameters. The technique in this paper is our first attempt to study and model this dependency between two major configuration parameters –e.g., number of mappers, and number of reducers– and total CPU usage in clock cycles of jobs in MapReduce environment. Briefly, our contributions in this paper are:

- Study the influence of configuration parameters on the performance of executing a job (here, total CPU usage in clock cycles) in MapReduce environments.
- Model this dependency using polynomial regression to predict total CPU usage in clock cycles of the same job on the same input data size.
- Briefly study the influence of input data scaling on total CPU usage in clock cycles of jobs.

These enable a user to choose suitable values for configuration parameters, improve the performance of executing his job, and predict the total CPU usage in clock cycles of his job on different input sizes. It is worth noting that because our provisioning model is focused on the overall performance of an application, it cannot provide detailed information regarding its internal steps –e.g., identifying parts of an application that are more CPU usage in clock cycles compared with its other parts. Moreover, complexity degree of an application along with a proper model selection can significantly influence accuracy of our model; thus, results are expected to be less accurate for highly complex applications. It should be noted that all realistic jobs selected for provision validation are moderate/high CPU intensive jobs. This is because analysing of total CPU usage in clock cycles is the most important factor in CPU intensive jobs; while for I/O jobs, I/O utilization should be studied.

## 2. Related Work

Early works on analysing/improving MapReduce performance started almost since 2005; such as an approach by Zaharia et al [7] that addressed problem of improving the performance of Hadoop for heterogeneous environments. Their approach was based on the critical assumption in Hadoop that only targets homogeneous cluster nodes; i.e., Hadoop assumes homogenous nodes to schedule its tasks and stragglers. A statistics-driven workload modelling was introduced in [9] to effectively evaluate design decisions in scaling, configuration and scheduling. Their framework was used to make practical suggestions to improve energy efficiency of MapReduce applications. Authors in [10] proposed a theoretical study on the MapReduce programming model which characterizes the features of mixed sequential and parallel processing in MapReduce.

Performance prediction in MapReduce has been another important issue. In [11], the variation effect of Map and Reduce slots on the performance has been studied. Also, it was observed that different MapReduce applications may result in different CPU and I/O patterns. Then a fingerprint based method is utilized to predict the performance of a new MapReduce application based on the studied applications. The idea of pattern matching was used in [12] to find the similarity between CPU time patters of a new application and applications in database. Then it was concluded that if two applications show high similarity for several setting of configuration parameters it is very likely their optimal values of configuration parameters also be the same. Authors in [5] also used historical execution traces of applications on MapReduce environment for profiling and performance modelling and prediction. A modelling method was proposed in [7] to predict the total execution time of a MapReduce application; they used Kernel Canonical Correlation Analysis to obtain the correlation between the performance feature vectors extracted from MapReduce job logs, and map time, reduce time, and total execution time. These features were acknowledged as critical characteristics for establishing any scheduling decisions. Authors in [13, 14] reported a basic model for MapReduce computation utilizations. Here, at first, the map and reduce phases were modelled using dynamic linear programming independently; then, these phases were combined to build a global optimal strategy for MapReduce scheduling and resource allocation. Another study in [8] proposed a resource provisioning framework to predict how much resources a user job needs to be completed by a certain time. This work also studied the impact of failures on the job completion time. To the best of our knowledge, most of resource provisioning methodologies in MapReduce environment address predicting of execution time of jobs; there is no specific research on studying (1) CPU, memory, and I/O cost of such jobs, and (2) dependency between configuration parameters of MapReduce environment and performance of execution of jobs (e.g., execution time, CPU usage in clock cycles). In sections 3.1 and 4.2, the importance of these two issues will be further explained.

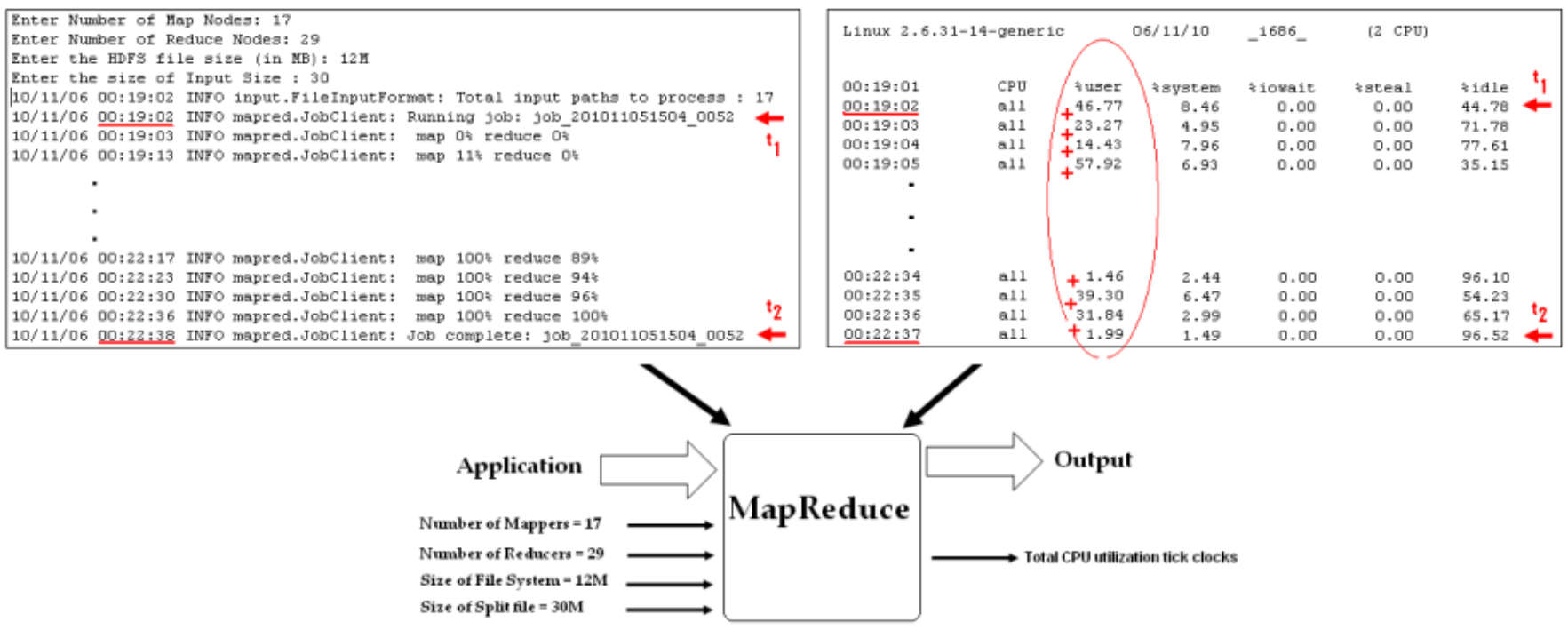

*Figure 2. the flow of the MapReduce job in Hadoop (left) and CPU usage time series extracted from actual system (right). This value is then converted to total CPU usage in clock cycle based on the platform's operating frequency.*

# 3. Application Modelling in MapReduce

In commercial clouds (such as Amazon EC2), the problem of allocating appropriate number of machines for a proper time frame strongly depends on an application; user is responsible to set these values properly [15]. Thus, to estimate how much resource (in CPU cost, I/O cost, and Memory cost) a job requires in total enables user to make educated decisions to hire appropriate number of machines. In MapReduce environments, this problem becomes more important as number of machines cannot be changed after starting a job.

## *3.1. Profiling total CPU usage in clock cycles*

For each application, we generate a set of jobs –i.e., an experiment of application on MapReduce environment– with different values of two MapReduce configuration parameters –i.e., number of mappers and reducers– on a given platform. While running each job, the CPU usage in clock cycles of the job is gathered –as training data– to build a trace for future deployments; such data can be easily gathered through functions provided in XenAPI with almost no overhead. Within the system, we sampled the CPU usage in clock cycles of a job, for each machine, from the time the mappers start to the time all reducers finish with time interval of one second as $\{C_{t_0}, C_{t_1}, \dots, C_{t_N}\}$ (figure 2-left). Then, total CPU usage in clock cycles of the job is calculated as (figure 2-right):

$$ncpu = \sum_{k=1}^{M} \left( \sum_{i=1}^{N} C_{t_i,k} \right) \times f_{clock,k}$$

where $M$ , $N$, and $f_{clock,k}$ are number of machines in cluster, number of CPU usage in clock cycles per seconds, and CPU clock frequency of k-th machine in cluster, respectively; for homogenous cluster , CPU clock frequency of all machines are the same.

Total CPU usage in clock cycles is an independent metric from number of machines in cluster. This means total CPU usage in clock cycles of a job should not significantly change on two clusters with different number of the same machines and configuration. For a cluster with $R$ machines, and a job with $T$ execution time, the following statements should be almost correct:

- A cluster with $\frac{R}{2}$ machines, the same configuration, and with the same CPU clock frequency should finish the job in $2T$ time.
- A cluster with $R$ machines, and the same configuration but half CPU clock frequency should finish the job at $2T$ time.

Total CPU usage in clock cycles on a job on the same clusters, however, can change for different values of configuration parameters – the purpose of this study.

### *3.2. Total CPU usage in clock cycles model using polynomial regression*

The next step is to create a model for an application on MapReduce environment by characterizing the relationship between a set of MapReduce configuration parameters and CPU usage in clock cycles metric. The problem of such a modeling –based on linear regression– involves choosing of suitable coefficients for the model to better approximate a real system response time [16, 17].

Consider the linear algebraic equations for $K$ different jobs of an application $(\varphi_i)$ with different sets of two configuration parameters values as follows:

$$\begin{cases} ncpu_{\varphi_i}^{(1)} = a_{0,\varphi_i} + a_{1,\varphi_i} M^{(1)} + a_{2,\varphi_i} (M^{(1)})^2 + a_{3,\varphi_i} R^{(1)} + a_{4,\varphi_i} (R^{(1)})^2 \\ ncpu_{\varphi_i}^{(2)} = a_{0,\varphi_i} + a_{1,\varphi_i} M^{(2)} + a_{2,\varphi_i} (M^{(2)})^2 + a_{3,\varphi_i} R^{(2)} + a_{4,\varphi_i} (R^{(2)})^2 \\ \vdots \\ ncpu_{\varphi_i}^{(k)} = a_{0,\varphi_i} + a_{1,\varphi_i} M^{(k)} + a_{2,\varphi_i} (M^{(k)})^2 + a_{3,\varphi_i} R^{(k)} + a_{4,\varphi_i} (R^{(k)})^2 \end{cases} \quad (1)$$

where $ncpu_{\varphi_i}^{(i)}$ is the actual value of total CPU usage in clock cycles of the application $\varphi_i$ in the $j^{th}$ job on MapReduce environment and $S^{(j)} = (M^{(j)}, R^{(j)})$ are the MapReduce configuration parameters; $M^{(j)}$ as the number of mappers, and $R^{(j)}$ as the number of reducers. Using the above definition, the approximation problem turns into estimating values of $\widehat{a_{0,\varphi_i}}, \widehat{a_{1,\varphi_i}}, \widehat{a_{2,\varphi_i}}, \widehat{a_{3,\varphi_i}}, \widehat{a_{4,\varphi_i}}$ to optimize a cost function between the approximation values and the actual values of total CPU usage in clock cycles. An approximated total CPU clock tick $(\widehat{ncpu_{\varphi_i}})$ of the application for an unseen job with configuration parameters $(M_*, R_*)$ is predicted as:

$$\widehat{ncpu_{\varphi_i}} = \widehat{a_{0,\varphi_i}} + \widehat{a_{1,\varphi_i}} M_* + \widehat{a_{2,\varphi_i}} M_*^2 + \widehat{a_{3,\varphi_i}} R_* + \widehat{a_{4,\varphi_i}} R_*^2 \quad (2)$$

There are a variety of well-known mathematical methods in the literature to calculate the variables $(\widehat{a_{0,\varphi_i}}, \widehat{a_{1,\varphi_i}}, \widehat{a_{2,\varphi_i}}, \widehat{a_{3,\varphi_i}}, \widehat{a_{4,\varphi_i}})$. One widely used in many application domains is the Least Square Regression which calculates the parameters in Eqn.2 by minimizing the following error:

$$error = \sqrt{\sum_{j=1}^{K} (\widehat{ncpu_{\varphi_i}^{(j)}} - ncpu_{\varphi_i}^{(j)})^2}$$

Least Square Regression theory claims that if:

$$H_{model} = \begin{bmatrix} 1 & M^{(1)} & (M^{(1)})^2 & R^{(1)} & (R^{(1)})^2 \\ 1 & M^{(2)} & (M^{(2)})^2 & R^{(2)} & (R^{(2)})^2 \\ & & \vdots & & \\ 1 & M^{(k)} & (M^{(k)})^2 & R^{(k)} & (R^{(k)})^2 \end{bmatrix}, H_{actual,\varphi_i} = \begin{bmatrix} ncpu_{\varphi_i}^{(1)} \\ ncpu_{\varphi_i}^{(2)} \\ \vdots \\ ncpu_{\varphi_i}^{(k)} \end{bmatrix},$$

$$A = \begin{bmatrix} \widehat{a_{0,\varphi_i}} \\ \widehat{a_{1,\varphi_i}} \\ \vdots \\ \widehat{a_{4,\varphi_i}} \end{bmatrix} \quad (3)$$

then the model satisfying the above error will be calculated as [17]:

$$A = (H_{model}^T H_{model})^{-1} H_{model}^T H_{actual,\varphi_i} \quad (4)$$

where $(.)^T$ denotes a transpose matrix. The set of configuration parameters values $\widehat{a_{0,\varphi_i}}, \widehat{a_{1,\varphi_i}}, \widehat{a_{2,\varphi_i}}, \widehat{a_{3,\varphi_i}}, \widehat{a_{4,\varphi_i}}$ is the model that approximately describes the relationship between total CPU usage in clock cycles of an application to two MapReduce configuration parameters.

## 4. Experimental Validation

In this section, we evaluate the effectiveness of our models using three realistic applications.

### *4.1. Experimental setting*

Three realistic applications are used to evaluate the effectiveness of our method. Our method has been implemented and evaluated on a private Cloud with the following specifications:

- Physical H/W: includes five servers, each one is an Intel Genuine with 3.00GHz clock, 1GB memory, 1GB cache and with 50GB of shared SAN hard disk.
- For virtualization, Xen cloud platform (XCP) has been used on top of the physical H/W. The XenAPI [18] provides functionality to directly manage virtual machines inside XCP. It provides binding in high level languages like Java, C# and Python. Using these bindings, it was possible to measure the performance of all virtual machines in a datacentre and live-migrate them.

- Virtual nodes are implemented on top of the XCP. The number of virtual nodes is chosen 20 with Linux image (Debian). The virtual nodes run Hadoop version 0.20.2 –i.e., Apache implementation of MapReduce developed in Java [19]. The XenAPI package is executed in background to monitor/extract the CPU utilization time series of applications (in the native system) [20]. For an experiment with a specific set of MapReduce configuration parameters values, statistics are gathered from "running job" stage to the "job completion" stage (arrows in Figure 2-left) with sampling time interval of one second. All CPU usages samples are then combined to form CPU utilization time series of an experiment.

In the training phase of our modelling, $64\,s$ets of jobs for each application are conduced where the number of mappers and reducers are integers with a value in $[4,8,12,16,20,24,28,32]$; the size of input data is fixed to $12G$. To overcome temporal changes, each job is repeated ten times. Then in the prediction phase, the accuracy of the application model is evaluated with 30 new/unseen jobs on the same input data size where the number of mappers and reducers are randomly selected from the integers $[4 \ldots 32]$.
Our benchmark applications are WordCount (used by leading researchers in Intel [21], IBM [6], MIT [22], and UC-Berkeley [7]), TeraSort (as a standard benchmark in the international TeraByte sort competition [23, 24] as well as many researchers in IBM [25, 26], Intel [21], INRIA [27] and UC-Berkeley [28]), and Exim Mainlog parsing [12, 29]. These benchmarks are used due to their striking differences as well as their popularity among MapReduce applications.

### *4.2 Evaluation Criteria*

We evaluate the accuracy of the fitted models, generated from regression based on a number of metrics [30]: Mean Absolute Percentage Error (MAPE), PRED(25) , Root Mean Squared Error (RMSE) and R2 Prediction Accuracy .

#### *4.2.1. Mean Absolute Percentage Error (MAPE)*

The Mean Absolute Percentage Error[30] for a prediction model is described as:

$$MAPE = \frac{\sum_{i=1}^{N} \frac{\left| {ncpu_{\varphi_k}}^{(i)} - {\widehat{ncpu_{\varphi_k}}}^{(i)} \right|}{{ncpu_{\varphi_i}}^{(i)}}}{N}$$

where ${ncpu_{\varphi_k}}^{(i)}$ is the actual total CPU usage in clock cycles of application $\varphi_k$, ${\widehat{ncpu_{\varphi_k}}}^{(i)}$ is the predicted total CPU usage in clock cycles and $N$ is the number of observations in the dataset. The smaller MAPE value indicates the better fit of the prediction model.

#### *4.2.2. PRED(25)*

The measure PRED(25)[30] is given as:

$$PRED(25) = \frac{\#\ of\ observations\ with\ relative\ error\ less\ than\ 25\%}{\#\ of\ total\ observations}$$

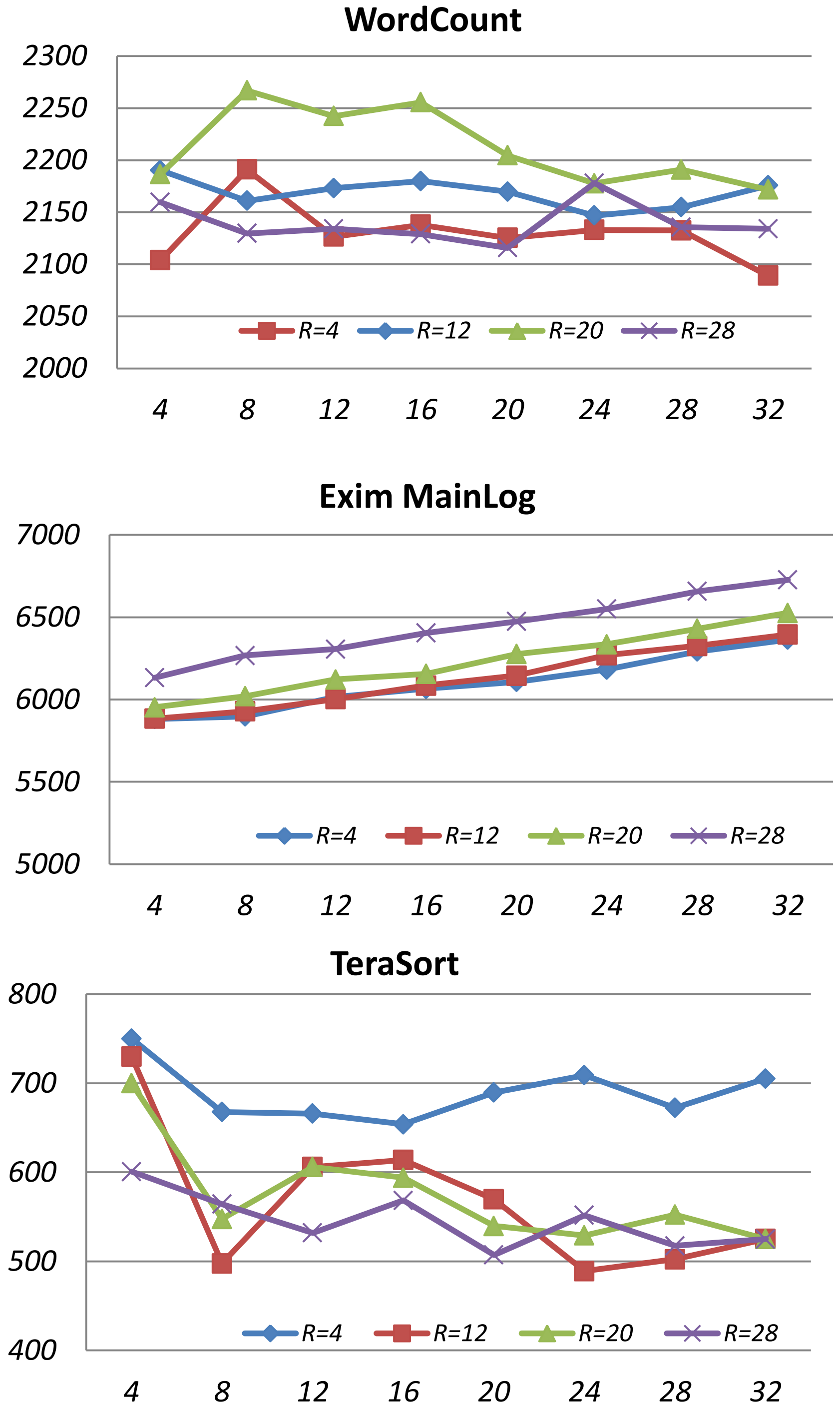

**(c)**

*Figure 3. The dependency between number of mappers and reducers and total CPU usage in clock cycle of jobs. The X-axis is number of mappers and Y-axis is total CPU usage(in tera clock cycle)*

It involves the percentage of observations whose prediction accuracy falls within 25% of the actual value. Closer value of PRED(25) to 1.0 implies a better fit of the prediction model.

*4.2.3. Root Mean Squared Error (RMSE)*

The metric Root Mean Square Error (RMSE)[30] is given by:

$$0 \leq RMSE = \sqrt{\frac{\sum_{i=1}^{N}({ncpu_{\varphi_k}}^{(i)} - \widehat{ncpu_{\varphi_k}}^{(i)})^2}{N}} \leq 1$$

More effective prediction results from smaller RMSE value.

*4.2.4. $\boldsymbol{R^2}$ Prediction Accuracy*

The $R^2$ Prediction Accuracy[30] − commonly applied to Linear Regression models as a measure of the goodness-of-fit of the prediction model– is calculated as:

$$0 \leq R^2 = 1 - \frac{\sum_{i=1}^{N}({ncpu_{\varphi_k}}^{(i)} - \widehat{ncpu_{\varphi_k}}^{(i)})^2}{\sum_{i=1}^{N}(\widehat{ncpu_{\varphi_k}}^{(i)} - \sum_{r=1}^{N}\frac{{ncpu_{\varphi_k}}^{(r)}}{N})} \leq 1$$

For a perfect prediction, $R^2 = 1.0$.

### *4.3. Results*

***Total CPU usage in clock cycles and configuration parameters:*** As mentioned earlier, there is a strong dependency between total CPU usage in clock cycles and the number of mappers and reducers in MapReduce environments. Figure 3 shows the dependency between these two configuration parameters and the total CPU usage in clock cycles for different jobs. One observation from this figure is that these applications behave differently when their number of mappers and reducers are increased. For example, Exim MainLog parsing shows a smooth linear relation, whereas such relation for WordCount and TeraSort is much more complicated. Another observation is that variations of the studied configuration parameters slightly change the difference between the highest and lowest total CPU usage in clock cycles for WordCount (9.5%) and Exim MianLog parsing (15.5%), while this difference for TeraSort is significant (50%).

***Prediction accuracy:*** To test the accuracy of the model, we use our proposed model to predict total CPU usage in clock cycles of several new/unseen jobs of an application with randomly set values of the two configuration parameters. We then ran the jobs on the real system and collect their total CPU in clock cycles to determine the prediction error. Figures 4 and 5 show the prediction accuracies and MAPE prediction errors; Table 1 reflects RMSD, MAPE, R2 prediction accuracy, and PRED for these

*TABLE 1. The prediction evalution*

| | **RMSD** | **MAPE** | **$R^2$ prediction accuracy** | **PRED** |
|---|---|---|---|---|
| **WordCount** | 0.208% | 1.59% | 0.851 | 1 |
| **Exim MainLog parsing** | 0.19% | 2.28% | 0.99 | 1 |
| **TeraSort** | 0.28% | 7.26% | 0.76 | 0.89 |

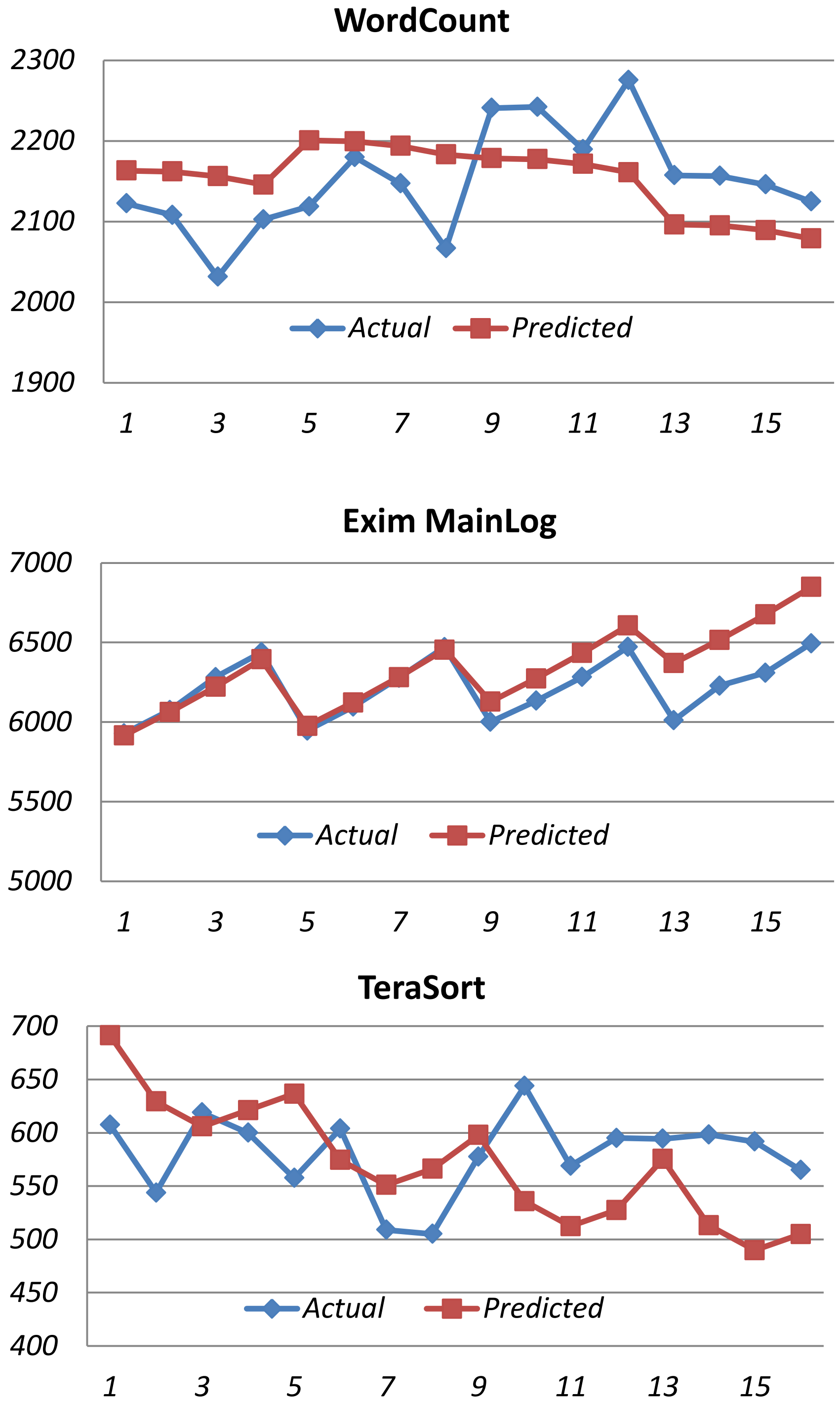

*Figure 4. The actual total CPU usage verses the model prediction for studied jobs. The X-axis is job ID and Y-axis is total CPU usage (in tera clock cycle)*

.
We found that most prediction evolution criteria are well satisfied for both WordCount and Exim MainLog parsing; it showed accuracy of our model. Although the MAPR value is still reasonable for TeraSort (~7%), it shows low values for other criteria. An educated guess to explain this phenomenon could be related to the significant difference between the highest and lowest total CPU usage in clock cycles of TeraSort (in figure 3), indicating that our modelling technique based on two-degree polynomial regression fails to correctly model the total CPU usage in clock cycles for TeraSort; therefore a better model must be used for this application.

***Input data scaling:*** figure 6 shows how total CPU usage in clock cycles of our three applications scales with increasing of input data size. As can be seen, there is a linear relation between these two metrics. Thus, total CPU usage in clock cycles modelling –calculated through the idea of this paper – for a fixed-size input data can be used for other data sizes as well.

### *4.3. Discussion and future work*

Although the obtained model can successfully predict the level of total CPU usage in clock cycles required for a few MapReduce applications, it shows some drawbacks. First, the total CPU usage in clock cycles of a job is modelled by averaging total CPU usage in clock cycles of the whole job from several traces. Many applications show quite different behaviour between their Map and Reduce phases: in some cases the Map is compute intensive, in others the Reduce or even both. Taking this into account, we would like to extend our model to a finer granular one. To this end, we like to split this model to cover CPU usage in clock cycles of both phases separately; i.e., instead of using a uniform average, we rather prefer to rely on a weighted average that emphasizes the CPU usage in clock cycles in each stage of the MapReduce computation. Second, the two successful applications –WordCount and Exim MainLog parsing– in our evaluation have almost linear complexity; and thus, their polynomial regression produced acceptable results. Our experiments also show that if such regression model is applied to programs with higher complexity (like TeraSort), their results are mostly unacceptable. To this end, we also like to consider other models –mostly non-linear regression– for more complex applications and provide a suit of regression techniques to cover almost all classes of applications.

## 6. Conclusion

In this work we proposed an accurate modelling technique to predict total CPU usage in clock cycles of jobs in MapReduce environment before their actual deployment on clusters and/or clouds. Such prediction can greatly help both application performance and effective resource utilization. To achieve this, we have presented an approach to model/profile total CPU usage in clock cycles of applications and applied polynomial regression model to identify correlation between two major MapReduce configuration parameters (number of mappers, and number of reducers) and the total CPU usage in clock cycles of an application. Our modelling technique can be used by both users/consumers (e.g., application developers) and service providers in the cloud for effective resource utilization. Evaluation results show that prediction error of total computation clock cycle of specific applications could be as less as 8%.

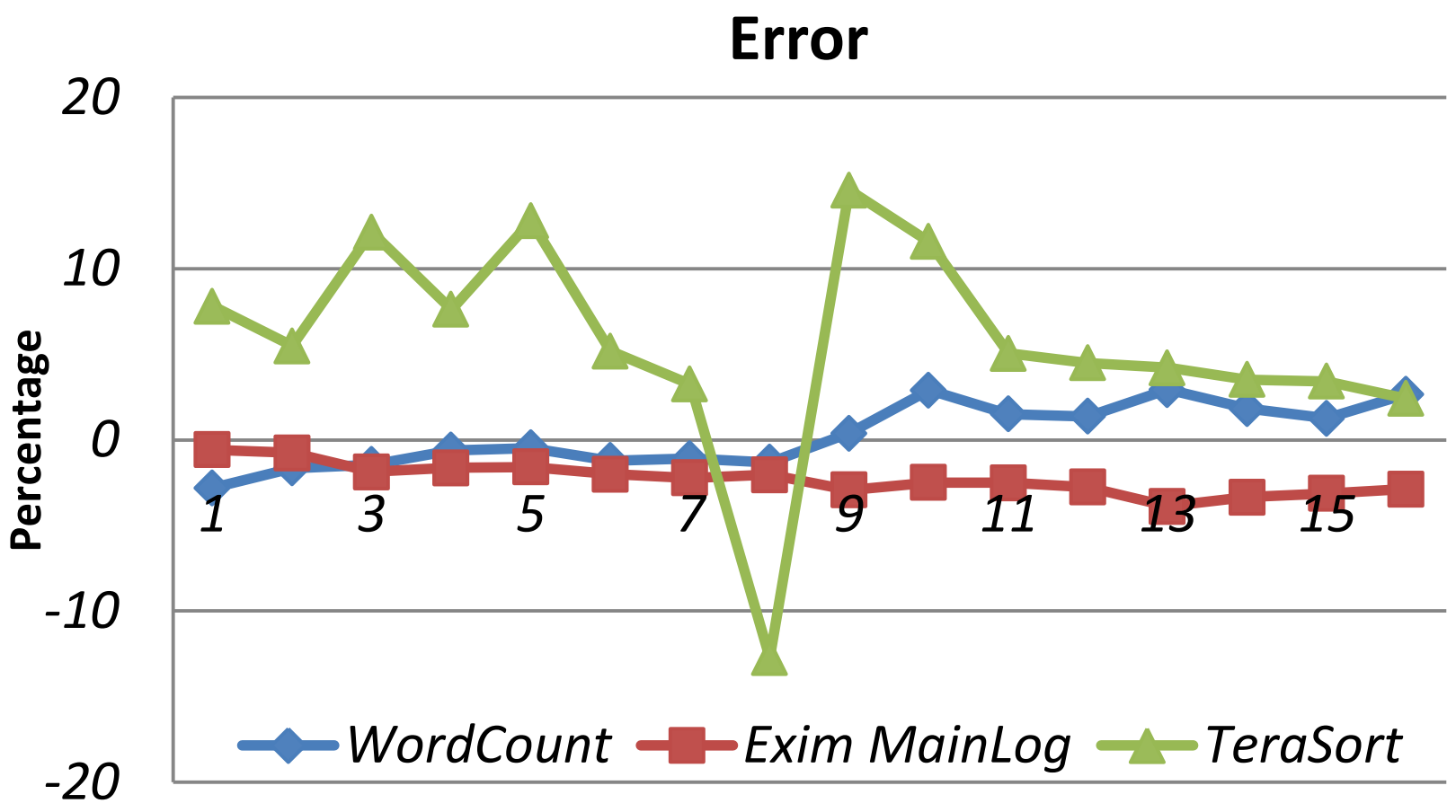

*Figure 5. The error between actual total CPU usage in clock cycle and the model prediction. The X-axis is job ID while Y-axis is percentage of error*

## Acknowledgment

Mr. N. Babaii Rizvandi's work is supported by National ICT Australia (NICTA). Professor A.Y. Zomaya's work is supported by an Australian Research Council Grant LP0884070.

## References

[1] *Hadoop example for Exim logs with Python*. Available: http://blog.gnucom.cc/2010/hadoop-example-for-exim-logs-with-python/

[2] N. B. Rizvandi, A. J. Boloori, N. Kamyabpour, and A. Zomaya, "MapReduce Implementation of Prestack Kirchhoff Time Migration (PKTM) on Seismic Data," presented at the The 12th International Conference on Parallel and Distributed Computing, Applications and Technologies (PDCAT), Gwangju, Korea 2011.

[3] K. Arumugam, Y. S. Tan, B. S. Lee, and R. Kanagasabai, "Cloud-enabling Sequence Alignment with Hadoop MapReduce: A Performance Analysis," presented at the 2012 4th International Conference on Bioinformatics and Biomedical Technology, 2012.

[4] *NCBI*. Available: http://www.ncbi.nlm.nih.gov/

[5] S. Kavulya, J. Tan, R. Gandhi, and P. Narasimhan, "An Analysis of Traces from a Production MapReduce Cluster," presented at the Proceedings of the 2010 10th IEEE/ACM International Conference on Cluster, Cloud and Grid Computing, 2010.

[6] K. Kambatla, A. Pathak, and H. Pucha, "Towards Optimizing Hadoop Provisioning in the Cloud," presented at the the 2009 conference on Hot topics in cloud computing, San Diego, California, 2009.

[7] M. Zaharia, A. Konwinski, A. D.Joseph, R. Katz, and I. Stoica, "Improving MapReduce Performance in Heterogeneous Environments," *8th USENIX Symposium on Operating Systems Design and Implementation (OSDI 2008),* pp. 29-42, 18 December 2008.

[8] A. Verma, L. Cherkasova, and R. H. Campbell, "Resource Provisioning Framework for MapReduce Jobs with Performance Goals," presented at the ACM/IFIP/USENIX 12th International Middleware Conference (Middleware), Lisbon, Portugal, 2011.

[9] Y. Chen, A. S. Ganapathi, A. Fox, R. H. Katz, and D. A. Patterson, "Statistical Workloads for Energy Efficient MapReduce," University of California at Berkeley,Technical Report No. UCB/EECS-2010-6, 2010.

[10] H. Karloff, S. Suri, and S. Vassilvitskii, "A model of computation for MapReduce," presented at the Proceedings of the Twenty-First Annual ACM-SIAM Symposium on Discrete Algorithms, Austin, Texas, 2010.

[11] K. Kambatla, A. Pathak, and H. Pucha, "Towards optimizing hadoop provisioning in the cloud," presented at the Proceedings of the 2009 conference on Hot topics in cloud computing, San Diego, California, 2009.

[12] N. B. Rizvandi, J. Taheri, A. Y. Zomaya, and R. Moraveji, "A Study on Using Uncertain Time Series Matching Algorithms in Map-Reduce Applications," *Concurrency and Computation: Practice and Experience,* 2012.

[13] A. Wieder, P. Bhatotia, A. Post, and R. Rodrigues, "Brief Announcement: Modelling MapReduce for Optimal Execution in the Cloud," presented at the Proceeding of the 29th ACM SIGACT-SIGOPS symposium on Principles of distributed computing, Zurich, Switzerland, 2010.

[14] A. Wieder, P. Bhatotia, A. Post, and R. Rodrigues, "Conductor: orchestrating the clouds," presented at the 4th International Workshop on Large Scale Distributed Systems and Middleware, Zurich, Switzerland, 2010.

[15] A.-m. Oprescu and T. Kielmann, "Bag-of-Tasks Scheduling under Budget Constraints " presented at the IEEE Second International Conference on Cloud Computing Technology and Science (CloudCom), Indianapolis, IN, U.S.A, 2010.

[16] T. Wood, L. Cherkasova, K. Ozonat, and a. P. Shenoy, "Profiling and Modeling Resource Usage of Virtualized Applications," presented at the Proceedings of the ACM/IFIP/USENIX 9th International Middleware Conference, Leuven, Belgium, 2006.

[17] N. B. Rizvandi, A. Nabavi, and S. Hessabi, "An Accurate Fir Approximation of Ideal Fractional Delay Filter with Complex Coefficients in Hilbert Space," *Journal of Circuits, Systems, and Computers,* vol. 14, pp. 497-506, 2005.

[18] D. Chisnall, *The definitive guide to the xen hypervisor.*, first ed.: Prentice Hall Press, 2007.

[19] *Hadoop-0.20.2*. Available: http://www.apache.org/dyn/closer.cgi/hadoop/core

[20] *Sysstat-9.1.6*. Available: http://perso.orange.fr/sebastien.godard/

[21] "Optimizing Hadoop Deployments," Intel Corporation2009.
[22] a. Mao, R. Morris, and M. F. Kaashoek, "Optimizing MapReduce for Multicore Architectures," Massachusetts Institute of Technology2010.
[23] S. Babu, "Towards automatic optimization of MapReduce programs," presented at the 1st ACM symposium on Cloud computing, Indianapolis, Indiana, USA, 2010.
[24] *Sort Benchmark Home Page*. Available: http://sortbenchmark.org/
[25] G. Wang, A. R. Butt, P. P, and K. Gupta, "A Simulation Approach to Evaluating Design Decisions in MapReduce Setups " presented at the MASCOTS, 2009.
[26] G. Wang, A. R. Butt, P. Pandey, and K. Gupta, "Using realistic simulation for performance analysis of mapreduce setups," presented at the Proceedings of the 1st ACM workshop on Large-Scale system and application performance, Garching, Germany, 2009.
[27] D. Moise, T.-T.-L. Trieu, L. Boug, #233, and G. Antoniu, "Optimizing intermediate data management in MapReduce computations," presented at the Proceedings of the First International Workshop on Cloud Computing Platforms, Salzburg, Austria, 2011.
[28] D. Gillick, A. Faria, and J. DeNero, "MapReduce: Distributed Computing for Machine Learning," www.icsi.berkeley.edu/~arlo/publications/gillick_cs262a_proj.pdf2008.
[29] N. B. Rizvandi, J. Taheri, and A. Y. Zomaya, "On using Pattern Matching Algorithms in MapReduce Applications," presented at the The 9th IEEE International Symposium on Parallel and Distributed Processing with Applications (ISPA), Busan, South Korea, 2011.
[30] S. Islam, J. Keung, K. Lee, and A. Liu, "Empirical prediction models for adaptive resource provisioning in the cloud," *Future Generation Comp. Syst.*, vol. 28, pp. 155-162, 2012.

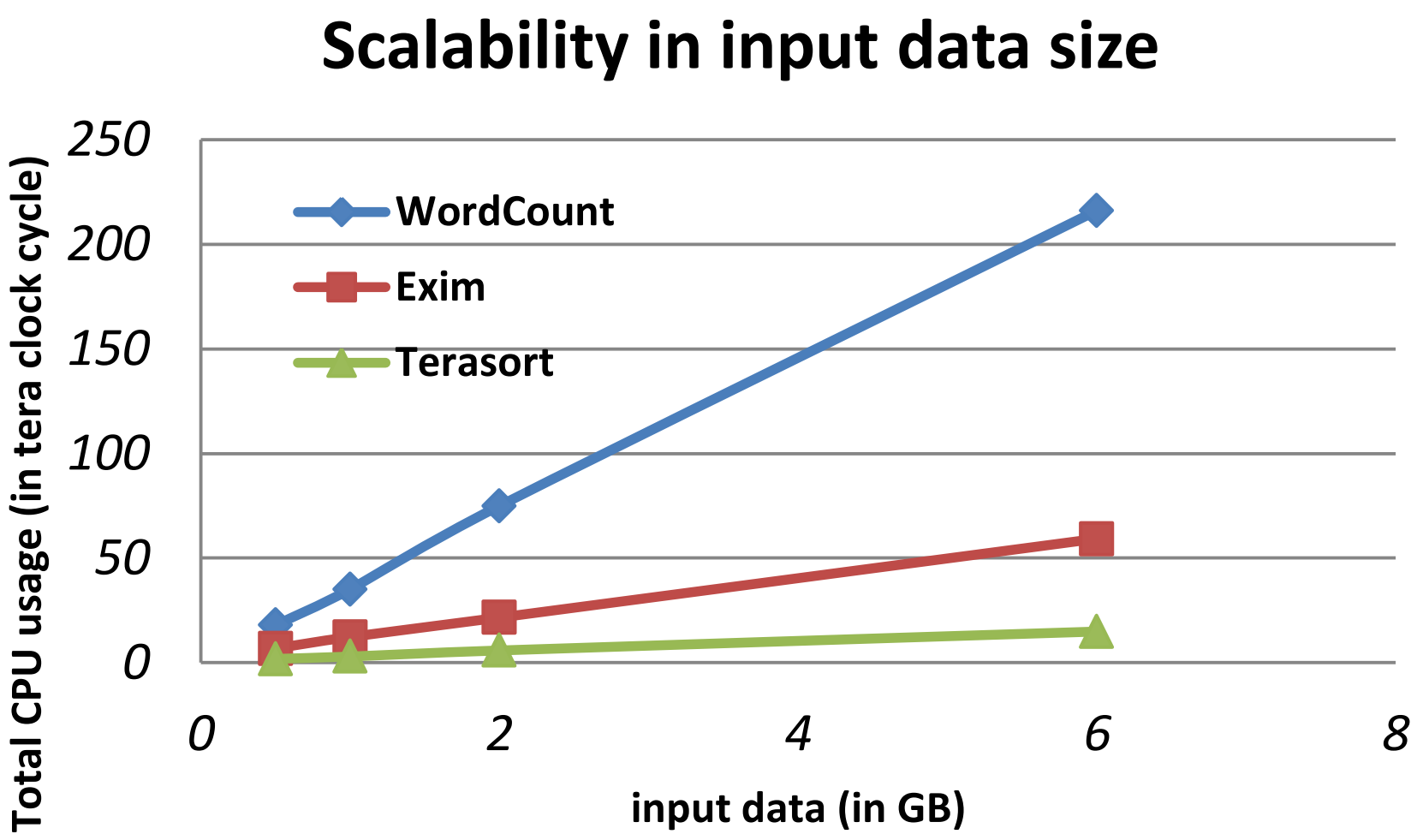

*Figure 6. total CPU usage in clock cycles and scalability in input data size*